\documentclass[conference, 10pt]{IEEEtran}
\IEEEoverridecommandlockouts
\usepackage{my-packages}
\usepackage{my-symbols}
\usepackage{my-typography}

\makeatletter

\begin{document}
\pagestyle{plain}
\title{An Asynchronous Distributed-Memory Parallel Algorithm for \kmer Counting}

\makeatletter
\newcommand{\newlineauthors}{%
  \end{@IEEEauthorhalign}\hfill\mbox{}\par
  \mbox{}\hfill\begin{@IEEEauthorhalign}
}
\makeatother

\author{
\IEEEauthorblockN{Souvadra Hati}
\IEEEauthorblockA{Email: souvadrahati@gatech.edu \\ Georgia Institute of Technology \\ Atlanta, GA, USA}
\and
\IEEEauthorblockN{Akihiro Hayashi}
\IEEEauthorblockA{Email: ahayashi@gatech.edu \\ Georgia Institute of Technology \\ Atlanta, GA, USA}
\and
\IEEEauthorblockN{Richard Vuduc}
\IEEEauthorblockA{Email: richie@cc.gatech.edu \\ Georgia Institute of Technology \\ Atlanta, GA, USA}
}

\maketitle
%%%%%%%%%%%%%%%%%%%%%%%%%%%%%%%%%%%%%%%%%%%%%%%%%%%%%%%%%%%%%%%%%%%%%%%

%------------------------------------------------------------
\begin{comment}
High-throughput DNA sequencing and analysis is cornerstone of modern life science and medicine.
% One of the fundamental kernel in DNA sequence analysis is \kc. 
\kc is the process of quantifying the frequency of length $k$ substrings in a DNA sequence.
It is a common kernel in numerous workloads in computational biology and can take up to 77\% of the total runtime of \textit{de novo} genome assembly.
The current \sota distributed memory \kc algorithm is based on the classic \BSP approach, and relies on multiple rounds of \mtom collectives, limiting the algorithm's performance.
We propose an asynchronous algorithm (\akc) that uses fine-grained, asynchronous messages to overcome the \BSP limitations while utilizing the network bandwidth efficiently.
\akc can perform strong scaling up to 256 nodes (512 sockets / 6K cores) and can count \kmers up to 9$\times$ faster than the \sota distributed-memory algorithm, and up to 100$\times$ faster than the shared-memory alternative.
We provide an analytical model to understand the hardware resource utilization of our asynchronous \kc algorithm and provide insights on the performance.
\end{comment}
\begin{abstract}
  This paper describes a new asynchronous algorithm and implementation for the problem of \kc, which concerns quantifying the frequency of length $k$ substrings in a DNA sequence.
  This operation is common to many computational biology workloads and can take up to 77\% of the total runtime of \textit{de novo} genome assembly.
  The performance and scalability of the current \sota distributed-memory \kc algorithm are hampered by multiple rounds of \mtom collectives.
  Therefore, we develop an asynchronous algorithm (\akc) that uses fine-grained, asynchronous messages to obviate most of this global communication while utilizing network bandwidth efficiently via custom message aggregation protocols.
  \akc can perform strong scaling up to 256 nodes (512 sockets / 6K cores) and can count \kmers up to 9$\times$ faster than the \sota distributed-memory algorithm, and up to 100$\times$ faster than the shared-memory alternative.
  We also provide an analytical model to understand the hardware resource utilization of our asynchronous \kc algorithm and provide insights on the performance.
  \end{abstract}
  

\acresetall

%------------------------------------------------------------
\section{Introduction} \label{intro}
\begin{figure}[ht!]
    \centering
    \begin{subfigure}[b]{0.48\textwidth}
        \centering
        \includesvg[width=\textwidth]{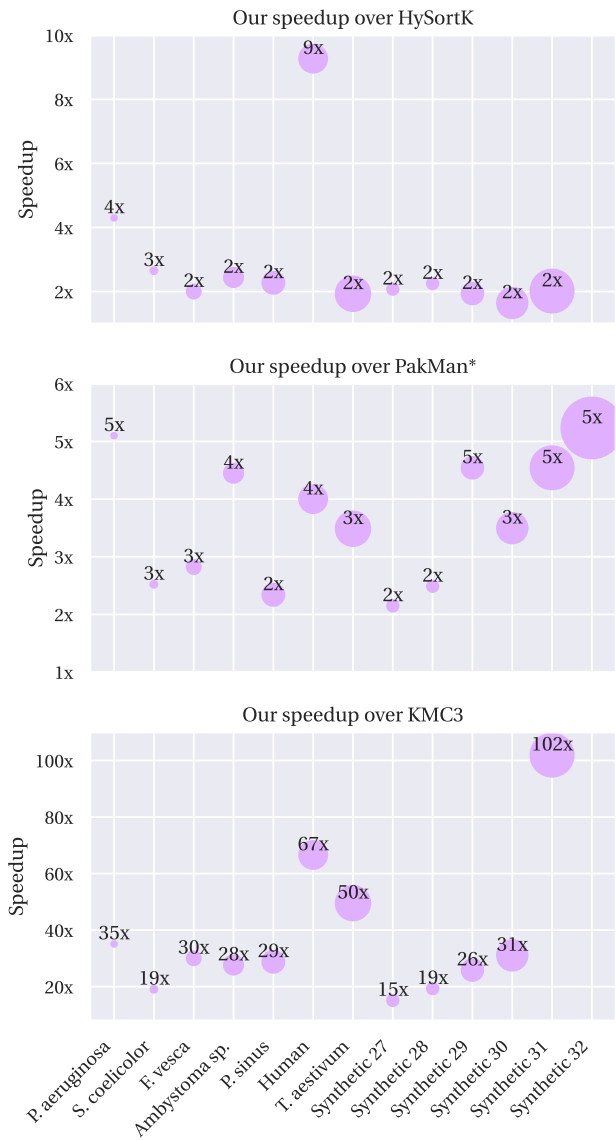}
    \end{subfigure}
    \caption{Speedup of \akc over baseline on synthetic and real genomes. Scatter dot sizes are proportional input size.}
    \label{fig:fig1}
\end{figure}
We consider the problem of \kc, which seeks to compute a histogram on \kmers, the string-based input values derived from a biological sequence (e.g., DNA, RNA, proteins; elaborated in \cref{problem-desc}).
\kc is a critical bottleneck in numerous computational genomics applications, including
\textit{de novo} genome assembly~\cite{cheng2021haplotype,chikhi2014informed,koren2018novo,pevzner2001eulerian,hipmer,pakman}, 
metagenome analysis~\cite{han2017concurrent,pellegrina2020fast,metahipmer,mhmgpu},
quality assessment of assembled genomes~\cite{rhie2020merqury}, 
error correction~\cite{kelley2010quake,salmela2017accurate}, 
repeat detection~\cite{li2003estimating}, 
RNA sequence analysis~\cite{audoux2017kupl} and 
cancer genomics~\cite{chong2017novobreak,khorsand2021nebula}, 
to name a few.

Myriad solutions exist for \kc on both shared memory and distributed memory systems, but these still motivate new approaches (\cref{intro-algo}).
Among shared memory methods~\cite{kmc2,kurtz_new_2008,li_mspkmercounter:_2015,li2018gpu,mamun_kcmbt_2016,marcais_fast_2011,pan2018optimizing,pandey2018squeakr,rizk_dsk:_2013}, consider the most popular and best in class implementation, KMC3~\cite{kmc3}.
Although it is widely used, its scalability is being stressed by the pace of growth of low-cost sequencing data generation:
on a relatively modest sequencing dataset consisting of 729 giga-base pairs (about 500 GB), a KMC3 demonstration run required nearly 2.5 hours and a minimum of 34 GB of RAM~\cite{kmc3}.
This limitation has motivated continued development of distributed-memory parallel alternatives with improved scaling for clusters or supercomputers~\cite{hysortk,kmerindsc18,hipmer,pakman,metahipmer,mhmgpu}.
But in one of the best of these, PakMan, \kmer counting takes up to 77\% of the total time for a short-read genome assembly pipeline~\cite{pakman};
and in another leading package, the GPU-accelerated metagenome assembly tool MetaHipMer2~\cite{mhmgpu}, about 50\% of the total runtime is spent on \kmer analysis.
Thus, we focus on further improving \kc in the distributed memory setting.

There are three primary challenges to scaling \kc:
1) poor data locality, which is intrinsic to the problem;
2) load imbalance, owing to the inherently skewed \kmer distributions that appear in real biological datasets and are only detectable at runtime;
and 3) the costs of synchronization.
The current \sota distributed memory \kc algorithm, HySortK~\cite{hysortk}, adopts a classic \BSP approach, relying on multiple rounds of \atoa collectives.
The resulting degree of synchronizations limits performance.

\textbf{Approach and contributions.}
We address the costs of synchronization by proposing a highly asynchronous alternative, which we refer to as \akc (\textbf{D}istributed \textbf{A}synchronous \textbf{k}-mer \textbf{C}ounter).
\akc requires just three (3) global synchronization steps, compared to a lower bound of two (2) and far fewer than HySortK, whose synchronization counts grow with input size (\cref{intro-algo}).
The ultimate results of our approach are summarized in \cref{fig:fig1}:
compared to a strong shared memory baseline (KMC3) and two distributed memory baselines (PakMan$^*$, HySortK), on various synthetic and real-world datasets, our approach offers 15--102$\times$ speedups over shared memory based KMC3, and 2--9$\times$ speedups over the distributed memory baselines.

Beyond the algorithmic approach, we contribute an implementation of \akc built on a recently proposed runtime, HClib Actor~\cite{jocs}, that targets distributed asynchronous programs (see \cref{comm-detail}).
This runtime is well-suited to algorithms like ours that are naturally expressed in a \FABSP style, which performs fine-grained asynchronous communication of small packets in between standard \BSP supersteps.
We also derive a simple analytical model of our \kc method (\cref{model-detail}) to help explain our empirical results (\cref{experiment}).
The model suggests that the performance achieved by our software is near optimal on our target machine, and also provides us insights regarding the use of hardware accelerators to improve performance.

% eof

%------------------------------------------------------------
% \section{Related Work} \label{related-work}
% \input{related}
%------------------------------------------------------------
\section{Background and Related Work} \label{problem-desc}
%\subsection{On counting \kmers}
%
% \kmers are substrings of length $k$ in a given set of strings.'
Given a finite alphabet $\Sigma$ and a fixed integer $k$, the set of \kmers, $\Sigma^k$, are all $k$-length strings that can be formed using $\Sigma$.
The task of \kc is to find the frequencies of all $\kappa \in \Sigma^k$ in a set of input strings.
In the case of DNA sequences, $\Sigma = \{A, C, G, T\}$, and the input is called `DNA reads', which are the output of sequencing machines.
% With the advent of high-throughput sequencing technologies, the amount of data generated by the sequencing experiments are massive, rendering the use of serial algorithm to count \kmers impractical.  
% This led to the development of parallel \kc algorithms that can count \kmers from massive sequencing datasets in reasonable time.
%
\subsection{Shared Memory Parallel Algorithms}
Jellyfish~\cite{marcais_fast_2011} was one of the earliest high performance \kmer counters. It used lock-free compare-and-swap atomics to count \kmers using a multithreaded hash table. 
Rizk et. al. proposed DSK~\cite{rizk_dsk:_2013}, a disk-based \kc algorithm that efficienctly utilized the disk I/O to count \kmers from datasets, too big to fit in the main memory of a node.
Deorowicz et. al.~\cite{kmc1, kmc2, kmc3} significantly expanded the ideas of DSK and designed KMC3, a faster and more efficient out of memory \kc algorithm, using minimizer-based~\cite{marccais2017improving} \kmer binning and multithreaded radixsort~\cite{raduls, raduls2}.
Since, not all \kmers are equally important in every genomics workloads, Melsted et. al. designed DFCounter~\cite{melsted_efficient_2011} that used Bloom filters to probabilistically avoid counting singleton \kmers and reduce the memory footprint.
Pandey et. al. expanded the idea of using probablistic data structures, and proposed a counting quotient filter~\cite{cqf_pandey} based approximate \kmer counter Squeakr~\cite{pandey2018squeakr}.
The reader is suggested to refer to the survery paper by Manekar et. al.~\cite{manekar2018benchmark} for a more comprehensive review of the shared-memory \kc algorithms.
Recent attempts have been made to offload \kc to \gpus~\cite{li2018gpu, mccoy2023singleton, nisa2021distributed, cheng2024rapidgkc} to leverage the additional computational power of the \gpus.
% However, we'll soon realize why GEMM accelerators, like \gpus are not the best choice of architecture for \kc. \textcolor{red}{(Change / Remove this line ? <-- Souvi)}
%
\subsection{Distributed Memory Parallel Algorithms}
The memory limitations posed by shared memory machines can be overcome by distributed memory \kc algorithms.
Such algorithms have three primary steps: 
(1) parse the input reads to generate the \kmers, 
(2) perform collective operations to distribute the \kmers among processors, and 
(3) get the final frequencies of the \kmers stored locally.
This strategy has been used in many distributed memory \kc module in popular genome and metagenome assembly tools like HipMer~\cite{georganas2014parallel, hipmer}, MetaHipMer~\cite{metahipmer}, PakMan~\cite{pakman}, and ELBA~\cite{elba}. 
The primary difference between these distributed memory \kc algorithms is the choice between hash table and sorting in the \textbf{third} step.
In 2018, Pan et. al. proposed a hash table based \kc algorithm~\cite{ kmerindsc18}, that built on top of a \kmer indexing tool KmerInd~\cite{KmerInd} and used AVX2 instructions to accelerate hash table query, and updates, to set the stage for the faster distributed memory \kmer counter.
Li and Guidi, in their 2024 article~\cite{hysortk} proposed HySortK, that surpassed the performance KmerInd by utilizing \OMP+ MPI based hybrid parallelism along with multithreaded radix-sorting. 
This makes HySortK the current \sota distributed memory \kc algorithm.
%------------------------------------------------------------
\begin{table}[b]
\centering
\caption{Algorithmic Symbols}
\begin{tabular}{ p{1.25cm} p{5cm}}
\toprule
%\multicolumn{2}{c}{Symbols} \\ \midrule
$\tau$    & Latency cost of remote communication \\
$\mu$     & Bandwidth cost of remote communication \\
$P$       & Number of processors \\
$n$       & Number of reads in input data \\
$m$       & Number of DNA/RNA bases in a read \\
$k$       & Length of a \kmer \\
$R[i][1:m]$       & The $i$-th input read, having $m$ bases \\
$b$       & Batch size for \BSP \kc \\
\bottomrule
\end{tabular}
\label{algoterm}
\end{table}
\section{Baseline + New Algorithms} \label{intro-algo}
We describe three algorithms:
a serial reference algorithm,
a baseline parallelization in a \BSP style,
and our new asynchronous algorithm in a \FABSP style.
The symbols used in our algorithms and analysis appear in \autoref{algoterm}.
On practical machines, $\tau \gg \mu$.

\subsection{Serial Algorithm}
Recall that \kc can be performed either using hash tables or by sorting.
We adopt the sorting-based approach since that is the current fastest distributed memory algorithm~\cite{hysortk} and used in the most popular shared-memory tool~\cite{kmc3}.

A serial sorting-based algorithm for \kc appears in \cref{algo:serial}.
The \texttt{Accumulate} function sweeps a sorted array of \kmers and counts the frequency of each \kmer.
The time complexity of \cref{algo:serial} is determined by the underlying sorting algorithm. 
A radixsort algorithm takes $\Theta{\left( mn \right)}$ time to sort \kmers from $n$ reads of $m$ DNA characters each.
\vspace{-11pt}
\begin{algorithm}
\caption{Serial Algorithm}
\label{algo:serial}
\small
\SetKwFunction{KCOUNT}{KmerCounting}
\SetKwFunction{GETFIRST}{GetFirstKmer}
\SetKwFunction{ENCODE}{Encode}
\SetKwFunction{ACC}{Accumulate}
\SetKwFunction{SORT}{Sort}
\SetAlgoLined
\KwData{$R$ (Set of reads), $k$ (\kmer length)}
\KwResult{$C$ $\leftarrow$ Ordered array of $\{$\kmer$,$ count$\}$}
\SetKwProg{Fn}{Function}{:}{}
\Fn{\KCOUNT{R, k}}{
    $T \leftarrow []$\;
    \For{$i \leftarrow 1 \text{ to } n$}{
        $kmer \leftarrow$ \GETFIRST{$R[i][1:k]$}\;
        $T.add(kmer)$\;
        \For{$j \leftarrow k+1 \TO m$}{
            $kmer \leftarrow (kmer \ll 2)$ \texttt{OR} \ENCODE{$R[i][j]$}\;
            $T.add(kmer)$\;
        }
    }
    \SORT{$T$}\;
    $C \leftarrow$ \ACC{$T$}\;
    \KwRet $C$\;
}
\vspace{1em} % Adds an empty line
\SetKwProg{Fn}{Function}{:}{}
\Fn{\GETFIRST{$R[1:k]$}}{
    $kmer \leftarrow 0$\;
    \For{$i \leftarrow 1 \TO k$}{
        $kmer \leftarrow (kmer \ll 2)$ \texttt{OR} \ENCODE{$R[i]$}\;    
    }
    \KwRet $kmer$\;
}
\end{algorithm}
\vspace{-15pt}
\subsection{BSP Algorithm}
The \BSP \kc algorithm extends algorithm~\ref{algo:serial} in three ways.
(1) Each distinct \kmer is owned by the \texttt{OwnerPE} that is responsible for counting it.
This convention ensures the local count of that \kmer in its `owner' processor is its final count.
(2) The \kc is done in batches of size $b$ to reduce the number of synchronizations required between the processors. 
The value of $b$ is user-tunable with typical values on current systems of $\approx 10^9$.
(3) The communication step is generally performed using \mtom collectives.
These ideas are embodied by \cref{algo:bsp}, which is implemented using \MPI blocking collectives as part of the \kc kernel of PakMan~\cite{pakman}.
HySortK extends this approach significantly by 
(1) incorporating MPI+OpenMP-based hybrid parallelism, which exploits the high core/socket structure of modern CPUs, and 
(2) using non-blocking collectives to increase computation and communication overlap.
% HySortK extends this approach by incorportating MPI+OpenMP based hybrid parallelism and non-blocking collectives to improve the scalabity of the algorithm, and added extra optimizations to improve the practical performance of the \BSP algorithm for real world datasets.  
%
\begin{algorithm}
\caption{\BSP Algorithm}
\label{algo:bsp}
\small
\SetAlgoLined
\SetKwFunction{KCOUNT}{KmerCounting}
\SetKwFunction{GETFIRST}{GetFirstKmer}
\SetKwFunction{ENCODE}{Encode}
\SetKwFunction{ACC}{Accumulate}
\SetKwFunction{SORT}{Sort}
\SetKwFunction{OWNER}{OwnerPE}
\SetKwFunction{FLUSHBUF}{FlushBuffer}
\SetKwFunction{MANYMANY}{\textcolor{red}{ManyToManyCollective}}
\KwData{$R$ (Set of reads), $k$ (\kmer length), $P$ (Processor count), $b$ (Batch size)}
\KwResult{$C$ $\leftarrow$ Ordered array of $\{$\kmer$,$ count$\}$}
\SetKwProg{Fn}{Function}{:}{}
\Fn{\KCOUNT{R, k, P, b}}{
    $T_s \leftarrow [[] \times P]$\;
    $T_r \leftarrow []$\;
    $N \leftarrow 0$\;
    \For{$i \leftarrow 1 \text{ to } n$}{
        $kmer \leftarrow$ \GETFIRST{$R[i][1:k]$}\;
        $p \leftarrow$ \OWNER{$kmer, P$}\;
        $T_s[p].add(kmer)$\;
        $N \leftarrow N + 1$\;
        \For{$j \leftarrow k+1 \TO m$}{
            $kmer \leftarrow (kmer \ll 2)$ \texttt{OR} \ENCODE{$R[i][j]$}\;
            $p \leftarrow$ \OWNER{$kmer, P$}\;
            $T_s[p].add(kmer)$\;
            $N \leftarrow N + 1$\;
            \If{$N = b$}{
                \FLUSHBUF{$T_s, T_r$}\;
                $N \leftarrow 0$\;
            }
        }
    }
    \FLUSHBUF{$T_s, T_r$}\;
    \SORT{$T_r$}\;
    $C \leftarrow$ \ACC{$T_r$}\;
    \KwRet $C$\;
}
\vspace{1em} % Adds an empty line
\SetKwProg{Fn}{Function}{:}{}
\Fn{\FLUSHBUF{$T_s, T_r$}}{
    $M \leftarrow []$\;
    \For{$i \leftarrow 1 \TO P$}{
        \ACC{$T_s[i]$}\;
        $M$.add($T_s$)\;
    }
    \MANYMANY{$M$}\;
    $T_r$.add($M$);
}
\end{algorithm}
\subsubsection*{Runtime analysis}
The runtime of the \BSP algorithm may be written as
\begin{align}
    T_{\text{BSP}} &= \tcomp + \left\lceil{\frac{mn}{bP}}\right\rceil \left(\tsync + \tcomm \right), \label{eq:bsp}
\end{align}
where $\tcomp$ is the local (per-process) computation time,
$\tsync$ is the time of one invocation of the collective primitive,
and $\tcomm$ is the time spent exchanging data.
%The computation time is dominated by the local sorting of \kmers. 
For a radixsort, the computation time is:
\begin{align}
    \tcomp  &= \Theta{\left( \frac{mn}{P} \right)}
\end{align}
We assume a tree-reduction algorithm to synchronize all the $P$ processors. 
We assume the best-case scenario for \mtom collective operation, where the runtime is the same as doing an \atoa collective.
\begin{align}
    \tsync  &= \Theta{\left( \tau \logp + \mu \logp \right)} \\
    \tcomm  &= \Theta{\left( \tau \logp + \mu bP \logp  \right)}
\end{align}
The final runtime of algorithm~\ref{algo:bsp}, after expanding the terms:
\begin{align}
    T_{\text{BSP}} &= \Theta \left({\frac{mn}{P}} + \tau \frac{mn}{bP} \logp + \mu mn \logp\right)
\end{align}
\subsection{\FABSP Algorithm (Our Algorithm)}\label{intro-algo:fabsp}
Algorithm~\ref{algo:fabsp} proposes an asynchronous alternative to the \BSP algorithm.
It specifically introduces the function \texttt{AsyncAdd}.
% We propose an asynchronous approach to parallel \kc by introducing the \texttt{AsyncAdd} function, as described in algorithm~\ref{algo:fabsp}.
This function represents a one-sided remote update, allowing the calling processor to asynchronously add a \kmer to the memory of the process (denoted by \texttt{OwnerPE(kmer)}) that owns the \kmer without direct involvement of the owner process.
The implementation details of this function are mentioned in \cref{comm-detail}.
%
% \vspace{-2pt}
\begin{algorithm}
\caption{\FABSP Algorithm}
\label{algo:fabsp}
\small
\SetAlgoLined
\SetKwFunction{KCOUNT}{KmerCounting}
\SetKwFunction{GETFIRST}{GetFirstKmer}
\SetKwFunction{ENCODE}{Encode}
\SetKwFunction{ACC}{Accumulate}
\SetKwFunction{SORT}{Sort}
\SetKwFunction{OWNER}{OwnerPE}
\SetKwFunction{FLUSHBUF}{FlushBuffer}
\SetKwFunction{ADD}{\textcolor{blue}{AsyncAdd}}
\SetKwFunction{BARRIER}{\textcolor{red}{GLOBAL BARRIER}}
\KwData{$R$ (Set of reads), $k$ (\kmer length), $P$ (Processor count), $b$ (Batch size)}
\KwResult{$C$ $\leftarrow$ Ordered array of $\{$\kmer$,$ count$\}$}
\SetKwProg{Fn}{Function}{:}{}
\Fn{\KCOUNT{R, k, P}}{
    $T \leftarrow []$\;
    \For{$i \leftarrow 1 \text{ to } n$}{
        $kmer \leftarrow$ \GETFIRST{$R[i][1:k]$}\;
        \ADD{$kmer$}\;
        \For{$j \leftarrow k+1 \TO m$}{
            $kmer \leftarrow (kmer \ll 2)$ \texttt{OR} \ENCODE{$R[i][j]$}\;
            \ADD{$kmer$}\;
        }
    }
    \BARRIER\textcolor{blue}{\;}
    \SORT{$T$}\;
    $C \leftarrow$ \ACC{$T$}\;
    \KwRet $C$\;
}
\end{algorithm}
% \vspace{-10pt}
%
\subsubsection*{Runtime analysis}
The time complexity of the algorithm~\ref{algo:fabsp} is similar to the algorithm~\ref{algo:bsp}, but with a single $T_{\text{sync}}$ term, instead of $\ceil{mn/bP}$ many required in the algorithm~\ref{algo:bsp}.
\begin{align}
    T_{\text{FABSP}} &= \Theta \left({\frac{mn}{P}} + \tau \logp + \mu mn \logp \right) \label{eq:fabsp}
\end{align}
From equations~\ref{eq:bsp} and~\ref{eq:fabsp}, we get:
\begin{align}
    T_{\text{BSP}} - T_{\text{FABSP}} &= \Theta\left(\tau \frac{mn}{bP} \log{P} \right) \label{eq:algodiff} \\
    \Rightarrow T_{\text{FABSP}} < T_{\text{BSP}}
\end{align}
In practice, we expect significant speedup over the \BSP algorithm because each round of synchronization causes CPU cycle waste, due to inherently skewed distribution of \kmers in complex genomes.
The \FABSP algorithm minimizes this issue using asynchronous execution. 
%  input-dependent load imbalance issues are also minimized in the asynchronous \FABSP algorithm.
%
% due to the input-dependent load imbalance among processors while processing real biological datasets.
% This load imbalance causes the \BSP algorithm to spend extra processor cycles just to wait for the processor with the highest workload to finish its assigned work, every time a synchronization is performed.
% The \FABSP algorithm minimizes this issue by allowing an asynchronous execution of the algorithm. 
% Hence, we expect to observe signficant speedup in the \FABSP algorithm over the \BSP counterpart despite both algorithm having similar time complexity in theory.
%------------------------------------------------------------
\section{Multilevel Aggregation of Communication} \label{comm-detail}
Our \FABSP algorithm (\cref{algo:fabsp}) uses one-sided, fine-grained messages, but making this run well in practice requires careful design.
While runtimes like MPI~\cite{mpi} and OpenSHMEM~\cite{openshmem} have native support for such operations via RDMA-based \texttt{Put} and \texttt{Get}, their direct use for smaller packets can be slow due to high latencies.
To hide these latencies, we use four layers of message aggregation protocols.
The runtime libraries we use (HClib+Actor, which builds on Conveyors) provide two of these layers (Layers 0 and 1), but then we add two more ``application-specific'' layers on top motivated by the needs of \kc.
\begin{table}[b]
    \centering
    \caption{Brief summary of different \conv protocols}
    \label{tab:convtable}
    \begin{tabular}{ c c c c }
    \toprule
    Protocol & Topology & Memory & \#Hops\\ \midrule
    $1D$ & All-Connected & $\mathbb{O}(P^2)$ & 1 \\
    $2D$ & 2D HyperX & $\mathbb{O}(P^{3/2})$ & 2 \\ 
    $3D$ & 3D HyperX & $\mathbb{O}(P^{4/3})$ & 3 \\
    \bottomrule
    \end{tabular}
    \vspace{5pt}
    \newline
    \noindent{The `Topology' here means the virtual topology that the processors follow to communicate, and not the physical topology of the interconnect.}
\end{table}
\subsection{Aggregation Layer 0 ($L_0$): \conv}
The lowest level of software aggregation is performed by \conv~\cite{conv}. 
It provides low-level APIs to store the data in a send-side buffer each time a \texttt{send} operation is called. 
Once the send-side buffer is full, the library invokes an RDMA-based \texttt{Put} to send the packets to the receive-buffer of the destination.
After the receive-buffer fills or the destination processor becomes idle, it goes through its received messages lazily and processes the packets.

There are three different modes, or protocols, used by the \conv library to perform scalable routing of messages among a large number of processors.
These are summarized in \cref{tab:convtable}. 
They trade-off extra buffer memory for reduced latency (measured by hops).
\subsection{Aggregation Layer 1 ($L_1$): HClib Actor Runtime}
The second level of aggregation is done by the runtime library, HClib\cite{jocs} which stores $C_1$ packets in each processor before sending them to the send-buffer of \conv.
($C_1$ is a tuning parameter.)
This extra layer of buffering ensures a seamless execution when the \conv buffers are full and/or busy being processed for communication. 
The runtime library is responsible for calling all the \conv APIs without user intervention, thereby hiding these aggregations from the application.
\subsection{Aggregation Layer 2 ($L_2$): Header Overhead}
For its 2D and 3D protocols, \conv adds a 32-bit header onto each packet to indicate the final destination. 
But \kmers of length $\leq 32$ are stored as 64-bit integers;
so, na\"ively adding them to the $L_1$ buffer incurs a header overhead that is $1/3$-rd of the data volume. 
To reduce that, each process maintains a buffer ($L_2$) to aggregate $C_2$ \kmers going to the same destination into a single packet before adding them to the $L_1$ buffers.
\subsection{Aggregation Layer 3 ($L_3$): The Curse of Complex Genomes}
Experimentally, the $L_0$ to $L_2$ aggregations work well for the majority of genomes.
% The first three layers of aggregation are enough to ensure fast asynchronous \kmer counting on the majority of genomes. 
However, complex mammalian and plant genomes often have few \kmers present in very high frequency (called heavy-hitters), thereby increasing load imbalance.
For example, the human genome is reported to have repeats of \texttt{(AATGG)$_n$} characters~\cite{hysortk}. 
This provides us with additional opportunities for more aggressive message aggregations to reduce the communication volume.
For this optimization, we need $2$ copies of the $L_2$ buffers, one called $L_2H$ (\texttt{HEAVY}-type), and $L_2N$ (\texttt{NORMAL}-type). 
To catch and treat the heavy-hitters, we first add the parsed \kmers in $L_3$ buffer. 
Once the $L_3$ buffer has $C_3$ elements in it, we can sort and accumulate on the $L_3$ buffer.
If the count of a \kmer is $> 2$, we send that \kmer as $\{kmer, count \}$ pair in the $L_2H$ buffer. 
Otherwise, we add the \kmer in the $L_2N$ buffer normally.
This reduces the communication volume for sending the \kmers in $L_3$ buffer to their destinations.
% This ensures minimal space usage for the \kmers inside the $L_3$ buffer, which in turn reduces the communication volume. 
The destination can detect the packet type (\texttt{HEAVY} vs. \texttt{NORMAL}) while processing it and can perform the final sort and accumulate accordingly.
\subsection{The full communication algorithm}
The complete algorithm for \textcolor{blue}{\texttt{AsyncAdd}}, including interactions among the aggregation protocol layers, appears in \cref{algo:fullcomm}.
\begin{algorithm}
\caption{\textcolor{blue}{\texttt{AsyncAdd}} Algorithm}
\label{algo:fullcomm}
\small
\SetAlgoLined
\SetKwFunction{KCOUNT}{KmerCounting}
\SetKwFunction{GETFIRST}{GetFirstKmer}
\SetKwFunction{ACC}{Accumulate}
\SetKwFunction{SORT}{Sort}
\SetKwFunction{OWNER}{OwnerPE}
\SetKwFunction{FLUSHBUF}{FlushBuffer}
\SetKwFunction{ADD}{\textcolor{blue}{AsyncAdd}}
\SetKwFunction{BARRIER}{\textcolor{red}{GLOBAL BARRIER}}
\SetKwFunction{PROCBUF}{\textcolor{violet}{ProcessReceiveBuffer}}
\SetKwFunction{ADLTHREE}{\textcolor{violet}{AddToL3Buffer}}
\SetKwFunction{ADLTWO}{\textcolor{violet}{AddToL2Buffer}}
\SetKwFunction{ADLONE}{\textcolor{violet}{AddToL1Buffer}}
\SetKwFunction{ADLZERO}{\textcolor{violet}{AddToL0Buffer}}
\SetKwFunction{EMP}{Empty}
\SetKwFunction{PUT}{\textcolor{red}{PUT}}

\SetKwProg{Fn}{Function}{:}{}

\Fn{\ADD{$kmer, T$}}{
    \ADLTHREE{kmer}\;
    \For{$p \in P$}{
        \If{$R_0[p].$size $= C_0$}{
            \PROCBUF{$T$}\;
        }
    }
}
\vspace{1em} % Adds an empty line
\Fn{\ADLTHREE{kmer}}{
    $L_3.$append$(kmer)$\;
    \If{$L_3$.size $= C_3$}{
        \SORT{$L_3$}\;
        \ACC{$L_3$}\;
        \For{$(k, \text{count}) \in L_3$}{
            \ADLTWO{k, \text{count}}\;
        }
    }
}
\vspace{1em} % Adds an empty line
\Fn{\ADLTWO{kmer, \text{count}}}{
    $p \leftarrow$ \OWNER{kmer}\;
    \If{count > 2}{
        $L_2H[p].$append$(kmer, count)$\;
        \If{$L_2H$.size $= C_2 / 2$}{
            \ADLONE{$L_2H[p], p$}\;
            \EMP{$L_2H[p]$}\;
        }
    }
    \Else{
        $L_2N[p].$append$(kmer)$\;
        \If{count $= 2$}{
            $L_2N[p].$append$(kmer)$\;
        }
        \If{$L_2N$.size $= C_2$}{
            \ADLONE{$L_2N[p], p$}\;
            \EMP{$L_2N[p]$}\;
        }
    }
}
\vspace{1em} % Adds an empty line
\Fn{\ADLONE{pkt, p}}{
    $L_1[p].$append$(pkt)$\;
    \If{$L_1[p].$size $= C_1$}{
        \ADLZERO{$L_1[p], p$}\;
        \EMP{$L_1[p]$}\;
    }
}
\vspace{1em} % Adds an empty line
\Fn{\ADLZERO{pktvec, p}}{
    $L_0[p].$concat$(pktvec)$\;
    \If{$L_0[p].$size $= C_0$}{
        \PUT{$L_0[p]$, R[MY$\_$PE], $p$}\;
        \EMP{$L_0[p]$}
    }
}
\vspace{1em} % Adds an empty line
\Fn{\PROCBUF{T}}{
    \For{$p \in P$}{
        \For{$e \in R[p]$}{
            \If{$(e0, e1) \in $ \texttt{HEAVY}}{
                $T.$append$(e0, e1)$
            }
            \Else{
                $T.$append$(e0,1)$\;
                $T.$append$(e1,1)$\;
            }
        }
        \EMP{$R[p]$}\;
    }
}
\tcp{$R$ is similar to $L_0$, but receives messages.}
\end{algorithm}

\begin{table}[b]
    \centering
    \caption{Aggregation Parameters}
    \label{tab:buffers}
    \begin{tabular}{ c c m{1.3cm} m{1.3cm} m{1.4cm} }
    \toprule
    Scope & Layer & Number of Buffers$/$PE & Element $/$Buffer & Memory $/$PE (Bytes) \\ \midrule
    Runtime     & $L_0$ & \centering{$\bluex$} & NA & $40K \times \bluex$ \\
    Runtime     & $L_1$ & \centering{$1$} & $C_1 = 1024$ & $264K$\\ 
    Application & $L_2$ & \centering{$P$} & $C_2 = 32$ & $ 264 \times P$ \\
    Application & $L_3$ & \centering{$1$} & $C_3 = 10K$ & $80K $ \\
    \bottomrule
    \end{tabular}
    \vspace{5pt}
    \newline
    \noindent{For $1D \Rightarrow x = 1$, $2D \Rightarrow x = 1/2$, and $3D \Rightarrow x = 1/3$.}
\end{table}

\subsection{Memory overhead of message aggregation}
% The four message aggregation protocols work together to reduce the communication time but pose extra memory overhead to buffer all the messages.
% Message aggregation makes the communication seamless and
The message aggregation memory overheads are summarized in \autoref{tab:buffers}, while that overhead compared to the \kc algorithm itself appears in \autoref{fig:total_mem}. 
% From this figure, we can easily convince ourselves that the extra memory usage is negligible for practical use cases. 
% Since the overall memory required increases with the number of processors, for the very high number of processors, weak scaling might be a challenge, due to the $O(P)$ per processor memory requirement of the $1D$ \conv buffers.
At high core counts, the $1D$ protocol memory becomes excessive, which can be mitigated by falling back to the $2D$ or $3D$ instead.

\begin{figure}[ht!]
    \centering
    \begin{subfigure}[b]{0.4\textwidth}
        \centering
        \includesvg[width=\textwidth]{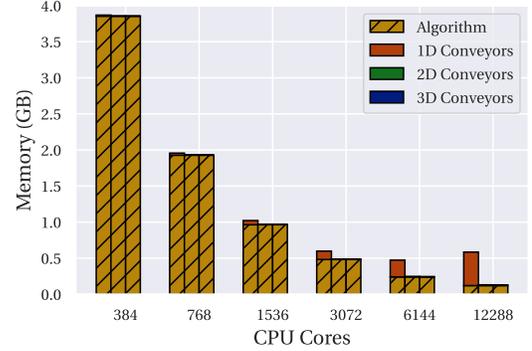}
    \end{subfigure}
    \caption{Per core memory overhead of 1D/2D/3D \conv for strong scaling experiment of \textit{Synthetic 32}.}
    \label{fig:total_mem}
\end{figure}
%

% \subsection{Choice of buffer parameters}

% The sizes of of $L_0$ and $L_1$ buffers, although tunable, fall under the scope of the runtime system. 
% Hence, we used the default $C_0$ and $C_1$ values. 

% The choice of $C_2$ depends on finding the smallest size where the overhead of the header will be negligible. 
% In our experiment we found $C_2 = 264$ Bytes to be that number.

% A naive way to determine the optimal size of $C_3$ buffer is to optimize the time it takes to sort and aggregate the $L_3$ buffer vs how much reduction in communication volume it can achieve. 
% If the value of $C_3$ is very high, that provides the algorithm more opportunities to find a lot of the high frequency \kmers and resulting in significant reduction of communication volume. 
% But, that comes at the extra cost of sorting a larger array. In our experiments we found $C_3 = 10000$ to be the optimal value on \Phoenix.  
%

%% Souvi: I don't think this picture is necessarily useful for the paper
% \begin{figure}[h!]
%     \centering
%     \begin{subfigure}[b]{0.48\textwidth}
%         \centering
%         \includesvg[width=\textwidth]{figures/buffer_overhead.svg}
%     \end{subfigure}
%     \caption{The distribution of memory among the four different aggregation layers.}
%     \label{fig:totel_mem}
% \end{figure}
%------------------------------------------------------------
%
\section{Analytical Model} \label{model-detail}
% \subsection{Analytical Model of asynchronous \kmer counting}
We propose a simple analytical model of \kmer counting to understand the behavior of our algorithm in real machines.
The underlying assumptions in our model are: 
(1) Input and output are perfectly load balanced.
(2) A $p$-core node has $100\%$ intranode parallel efficiency.
(3) Algorithms are oblivious to the processor's cache hierarchy. 
(4) Processors have a two-level memory hierarchy, with an infinite main memory, a $Z$ bytes cache size, and $L$ bytes line size, with optimal line replacement policy.
The full \kmer counting workload can be decomposed into two phases:
\subsubsection*{\textbf{Phase 1:} \kmer generation and reshuffling}
In this phase, the algorithm parses the input reads and generates output \kmers which are then sent to their destination processors.

In the input FASTA/Q files, each DNA character is represented using an $8$-bit ASCII character.
The \kmer counting algorithm first converts the ASCII characters into a $2$-bit DNA encoding.
Then, the algorithm parses the input from start to finish and concatenates $k$ consecutive characters to build a \kmer.
From a single input read of $m$ DNA characters, we can generate $\kmernum$ \kmers. 
For faster computation, a \kmer of length $k$ is stored using $\kmerbits$-bit unsigned integer.
From this, we can easily deduce the computation time in phase 1, as: 
\begin{align}
    \tcomp^1 &= \frac{\kmernum}{P \cnode} \label{eq:t1comp}
\end{align}
Parsing the input reads results in $\inputcache$ cache misses and storing the \kmers in another array results in $\kmercache$ cache misses. 
This results in the intranode communication time as:
\begin{align}
    \tintra^1  &= \left[ \inputcache + \kmercache \right] \frac{L}{\bmem} \label{eq:t1intra}
\end{align}
After \kmer generation, each node sends the \kmers to their destination / `owner' nodes. 
This section resembles a \mtom communication pattern, where each node sends approximately $\kmernum \kmerbits / 8P$ Bytes of data to all other nodes. 
This results in $\kmernum \kmerbits / 4P$ Bytes of data transferred through the \nic of each node. 
Hence, on a fully connected network with combined bidirectional link bandwidth of $\blink$, the time to perform internode communication is:
\begin{align}
    \tinter^1 &= \frac{\kmernum \kmerbits}{4P \blink} \label{eq:t1inter}
\end{align}

%% Table dictating different values
\begin{table}[H]
\centering
\caption{Model parameters for \Phoenix}
\begin{tabular}{ p{2.5cm} p{0.7cm} p{1.5cm} }
\toprule
Parameter            &                       & Intel Node \\ \midrule
Peak INT64          & $\cnode$              & $121.9$ GOp/s    \\
Memory Bandwidth    & $\bmem$               & $46.9$ GB/s   \\ 
Fast Memory         & $Z$                   & $38$ MB       \\ 
Cacheline size      & $L$                   & $64$ B        \\
Link Bandwidth      & $\beta_{\text{link}}$ & $12.5$ GB/s   \\
\bottomrule
\end{tabular}
\label{model_tab}
\end{table}

\subsubsection*{\textbf{Phase 2:} Sorting and Accumulation}
In this phase, the received \kmers are first sorted and then accumulated to store them as a sorted array of $\{$\kmer,~count$\}$ pairs.

The computation time in this phase is dominated by the sorting algorithm. 
We use a hybrid sorting algorithm \cite{skasort} that starts with an in-place radix sort and falls back to comparison-based sorting using a heuristic. 
In our model, we assume the worst-case behavior of an in-place radix sort algorithm, where the algorithm parses through the data one byte at a time.
That will result in worst-case $\kmerbytes$ number of passes through the data to sort it completely.
This results in a computation time: 
\begin{align}
    \tcomp^2 &= \frac{\kmernum \kmerbits}{8 P \cnode} \label{eq:t2comp}
\end{align}
%
% For modeling the intranode communication, we assume the comparison based sorting used is cache optimal `funnelsort' \cite{cache1999}.
% Number of cache misses to sort $x$ items using funnelsort is $ \Theta{\left( 1 + (x/L)(1 + \max\{\log_{Z}{x}, 1\}  \right)}$. 
% Then, we estimate the overall cache misses in our hybrid algorithm as the minimum of a radix sort and funnelsort.
Similarly, the intranode communication time is:
\begin{align}
    \tintra^2 &= \left[ {\left( 1 + \frac{\kmernum \kmerbits}{8 P L} \right) \kmerbytes } \right] \frac{L}{\bmem} \label{eq:t2intra}
    % [ \ N &\coloneqq \kmernum \kmerbits / 8P \ ] \notag 
\end{align}
% 
% The term inside square brackets in equation~\ref{eq:t2intra} refers to the number of cache misses in phase $2$ of \kmer counting. 
% The constant term, $A$ is used to account for the multi-layer cache hierarchy of a modern processor.
% Using hardware counters on a Xeon processor, we found $A \approx 5.7$. 
%
\subsubsection{Total Costs}
The total communication time in phase 1 is either the sum or maximum of intranode and internode communication time as shown below:
\begin{align}
    \tcomm^1 &= \tintra^1 + \tinter^1 \ \ \text{ or } \label{eq:t1sum} \\
    \tcomm^1 &= \max{\left(\tintra^1, \tinter^1 \right)} \label{eq:t1max}
\end{align}
We built both models using~\autoref{eq:t1sum} and~\autoref{eq:t1max}. We call them the `Sum' and `Max' model respectively. 

The total time spent in phase 1 can be represented as:
\begin{align}
    T_{1} &= \max{\left( \tcomp^1 , \tcomm^1 \right)}
\end{align}
Similarly, the time to execute the phase 2 is:
\begin{align}
    T_2 &= \max{\left( \tcomp^2, \tintra^2 \right)}
\end{align}
\akc algorithm requires a \textit{GLOBAL BARRIER} between the two phases.
This makes it impossible to overlap both phases. 
Hence, the total time of the full algorithm is:
\begin{align}
    T_{\text{total}} &= T_1 + T_2
\end{align}
\subsection{Model Validation}
We use the parameter values from~\autoref{model_tab} for our analytical model.
The parameters are based on \Phoenix at \gt.
The $\cnode$ and $\bmem$ values are obtained using our microbenchmarks.
The $\cnode$ parameter tells us the maximum number of $64$-bit integer additions a single node of \Phoenix can perform.

We validated our analytical model against measured last-level cache misses reported by PAPI~\cite{papi}. % counters to measure the number of last-level cache misses in \Phoenix nodes in both phases of \kmer counting. 
\autoref{fig:hwcounts} demonstrates the prediction of our model and the observed last-level cache misses in our experiments.
The cache misses predicted by our model in phase 1 are slightly lower than the measured counts, which is expected since the model assumes a perfect cache replacement policy. 

%\textcolor{red}{[Souvi: Explain what's happening for smaller datasets. I don't have a real answer RIP ME !!]}
In the second phase, the predicted cache misses are based on the worst-case behavior of a radix sort algorithm. But, in reality, the sorting algorithm used can detect partially sorted arrays and skip sorting them, resulting in lower cache misses compared to our prediction. 
Even then, this effect is rather small once the data size is large enough.
\begin{figure}[h!]
    \centering
    \begin{subfigure}[b]{0.48\textwidth}
        \centering
        \includesvg[width=\textwidth]{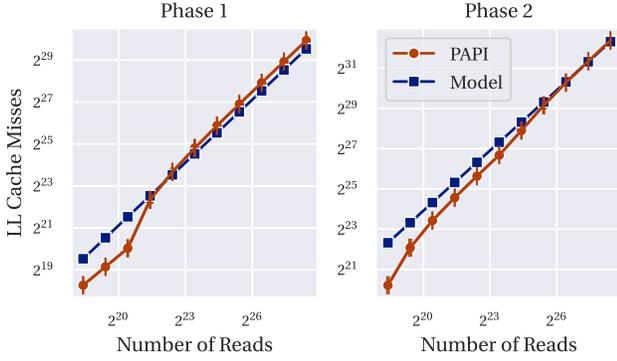}
    \end{subfigure}
    \caption{
        Last-level cache misses predicted by our analytical model and the observed values from the hardware counters. 
        We performed the experiments using $8$ nodes ($192$ cores) of \Phoenix. 
        Each experimental data point is the average of three consecutive runs and the error bars represent the standard deviation across runs.
    }
    \label{fig:hwcounts}
\end{figure}

\autoref{fig:modeltime} shows the predicted execution time of both phases of \kmer counting by our model with the experimentally observed numbers. 
In both cases, our analytical model underestimates the execution time but remains in the same ballpark as the actual experimental results. 
\begin{figure}[h!]
    \centering
    \begin{subfigure}[b]{0.48\textwidth}
        \centering
        \includesvg[width=\textwidth]{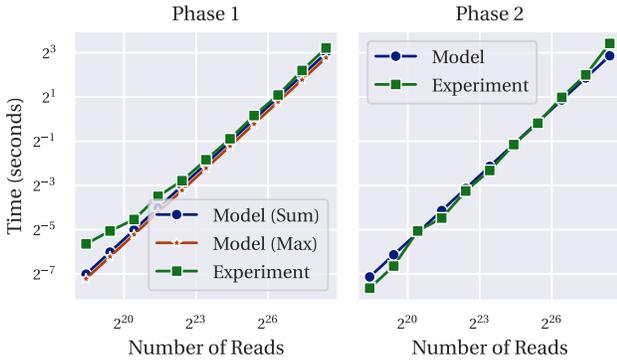}
    \end{subfigure}
    \caption{
        Time is taken by two phases of \kmer counting as measured in experiments and predicted by our model. 
        We performed the experiments using $8$ nodes ($192$ cores) of \Phoenix.
        Each experimental data point is the best observed time from three consecutive runs.
        }
    \label{fig:modeltime}
\end{figure}
\subsection{Insights from the analytical model}
\subsubsection*{Hardware resources utilization in \kmer counting}
We estimate the time to perform \kmer counting of \textit{Synthetic 30} dataset on $32$ nodes ($768$ cores) of \Phoenix, using our analytical model. 
\autoref{fig:distribution} summarizes the estimation of the model. 
We can observe that the time spent on computation is very small. 
The intranode and internode communication time takes up the majority of the total time making this workload purely bounded by how fast data can be moved either from the memory to the processor or between processors.
%
% \begin{figure}[h!]
%     \centering
%     \begin{subfigure}[b]{0.40\textwidth}
%         \centering
%         \includesvg[width=\textwidth]{figures/hwpie.svg}
%     \end{subfigure}
%     \caption{Percentage of total execution time spent in computation, intranode and internode communication in distributed \kmer counting of \texttt{Synthetic 30} dataset, using $32$ nodes ($768$ cores) as per our analytical model. We assume no computational communication overlap for this figure.}
%     \label{fig:distribution}
% \end{figure}
%
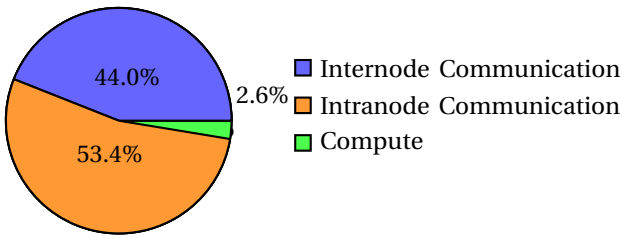
\begin{figure}[h!]
    \centering
    \begin{tikzpicture}
        \pie
            [text=legend, radius=1.5, color={blue!60, orange!80, green!70}]
            {44.0/Internode Communication, 53.4/Intranode Communication, 2.6/Compute}

        \pie
            [hide number, radius=1.5, color={blue!60, orange!80, green!70}]
            {2.6/{$\!\!2.6\%$}, 44.0/, 53.4/}

        \pie
            [radius=1.5, color={blue!60, orange!80, green!70}]
            {44.0/ , 53.4/}
    \end{tikzpicture}
    \caption{Percentage of total execution time spent in computation, intranode and internode communication in distributed \kmer counting of \textit{Synthetic 30} dataset, using $32$ nodes ($768$ cores) as per our analytical model. We assume no computational communication overlap for this figure.}
    \label{fig:distribution}
\end{figure}
\section{Experiments and Analysis} \label{experiment}
\begin{figure}
    \centering
    \includesvg[width=\linewidth]{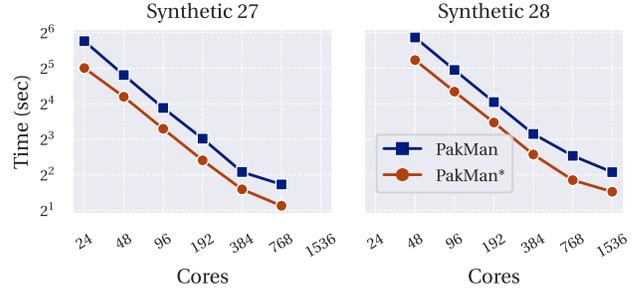}
    \caption{Use of radixsort in \MPI based PakMan (we call it PakMan*) results $2\times$ faster runtime.}
    \label{fig:pkmstar}
\end{figure}
\begin{figure*}[ht!]
    \centering
     \begin{subfigure}[b]{0.95\textwidth}
        \centering
        \includesvg[width=\textwidth]{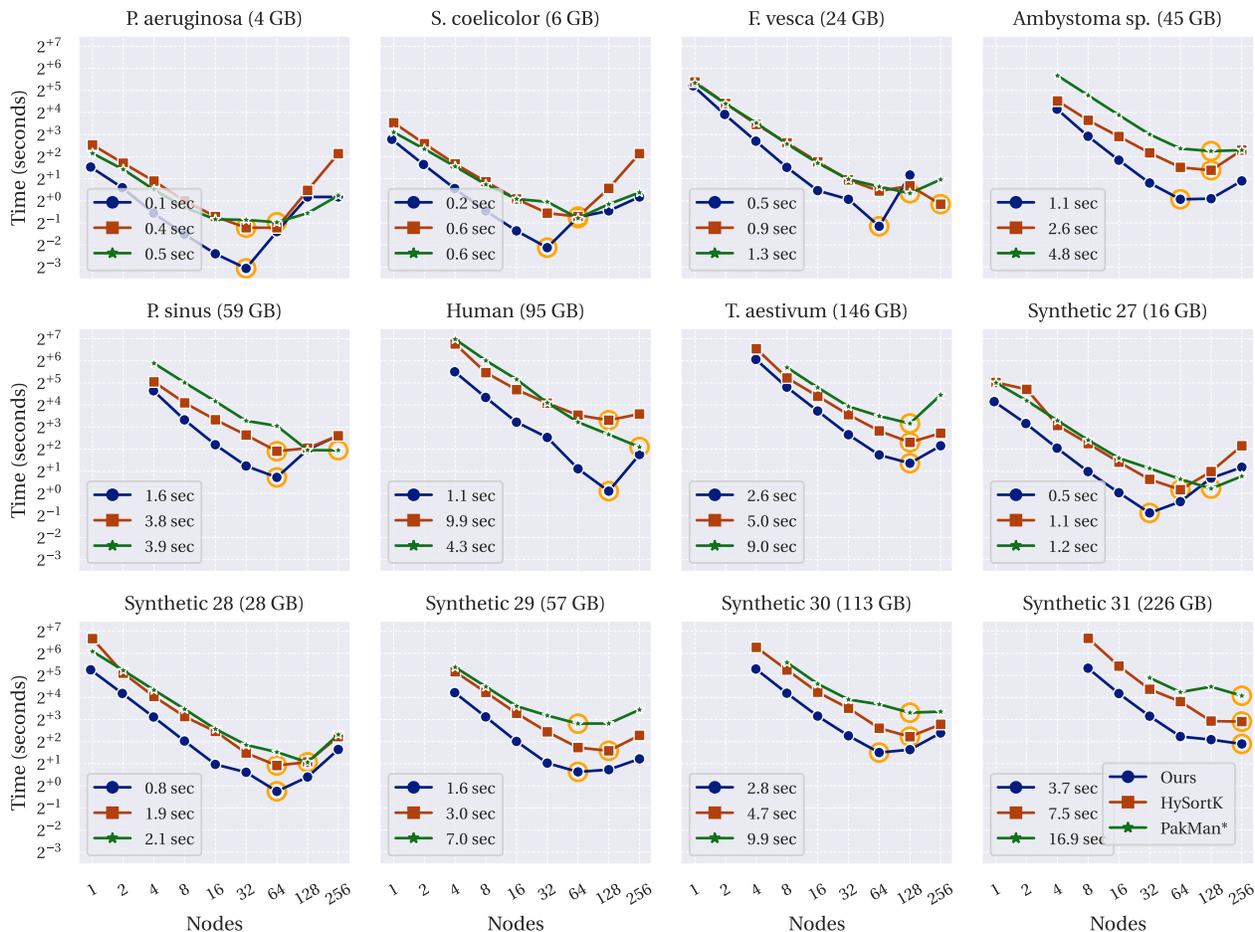}
    \end{subfigure}
    \caption{Strong scaling on synthetic and real genomes using up to $256$ nodes / $6144$ cores. We only use $L_3$ aggregation protocol on \textit{Human} and \textit{T. aestivum}, because they are known in the literature to have high-frequency \kmers.}
    \label{fig:strongscale}
\end{figure*}
\begin{figure}[h!]
    \centering 
    \begin{subfigure}[b]{0.42\textwidth}
        \centering
        \includesvg[width=\textwidth]{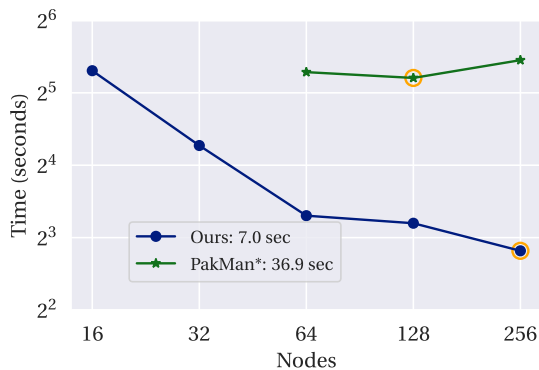}
    \end{subfigure}
    \caption{Strong scaling on our largest dataset, \textit{Synthetic 32} (451 GB). PakMan* gave OOM error for $16$ \& $32$ nodes. HysortK did not run for any configuration.}
    \label{fig:ss32}
\end{figure}

%\subsection{Experimental Setup}
We conducted experiments using \Phoenix cluster at \gt. 
At the time of writing, \Phoenix has $453$ Intel nodes and $8$ AMD nodes connected via Infiniband 100HDR interconnect. 
The Intel nodes have dual-socket Xeon Gold 6226 CPUs clocked at $2.7$~GHz, with $24$~cores and $192$~GB DDR4-2933~MHz DRAM memory. 
The AMD nodes have dual-socket EPYC~7742 CPUs clocked at 2~GHz, with $128$~cores and $512$~GB DDR4 DRAM memory. 
% The input datasets are stored on the Lustre filesystem for efficient parallel I/O. 
We use $256$ Intel nodes for distributed memory experiments and individual AMD and Intel nodes for shared memory experiments.
We use \textsf{python/3.10.10}, \textsf{gcc/12.3.0}, \textsf{openmpi/4.1.5}, and  \textsf{openshmem 1.4}.
\akc code is available as open-source software at \url{https://github.com/Souvadra/dakc/}.

In distributed memory, we exclude I/O time since it is out of scope of this work. 
For shared memory experiments, we mention the total time including I/O because KMC3's output log combines I/O and compute.
HySortK's I/O is poorly optimized. 
Hence, we use \akc's I/O time as the best-case scenario for HySortK, making it's total time as strong as possible.
Each algorithm is tasked with counting \kmers for $k = 31$ from count $= 1$ to the maximum supported count. 
We report the best of $3$ consecutive runs. 
% We choose $k=31$ in the experiments because it is the largest odd $k$-mer that can be represented using $64$-bit integers.
%

%\subsection{Dataset Used}
%
\autoref{datasets} summarizes the synthetic and real datasets used in our experiments.
For synthetic datasets, we generate input files in the standard FASTQ format using ART Illumina Simulator~\cite{huang2012art} on a synthetic genome, sampled uniformly randomly from the alphabet $\Sigma = \{A,C,G,T\}$.
\textit{Synethetic XY} refers to the FASTQ file generated from a genome, $2^{XY}$ DNA bases long.
The real datasets are downloaded from the NCBI SRA database~\cite{ncbi} and are converted to FASTQ format using the \textsf{fasterq-dump} tool from the SRA toolkit~\cite{sratoolkit}.
We only use the first of the two paired-end reads. % for our experiments.  

\begin{table}
\caption{Datasets Used in Experiments}
\begin{tabular}{l l l l l}\toprule
Data & Reads & Read & Fastq Size & Name \\
     &       & Length      &            &      \\
\midrule
Synthetic 20 & 349,500         & 150   & 0.11 MB  & -   \\ 
Synthetic 21 & 699,050         & 150   & 0.22 MB  & -   \\
Synthetic 22 & 1,398,100       & 150   & 0.44 MB  & -   \\
Synthetic 23 & 2,796,200       & 150   & 0.9 GB   & -   \\ 
Synthetic 24 & 5,592,400       & 150   & 1.8 GB   & -   \\ 
Synthetic 25 & 11,184,800      & 150   & 3.5 GB   & -   \\ 
Synthetic 26 & 22,369,600      & 150   & 7.0 GB   & -   \\
Synthetic 27 & 44,739,200      & 150   & 16.0 GB  & -   \\
Synthetic 28 & 89,478,450      & 150   & 28.0 GB  & -   \\
Synthetic 29 & 178,956,950     & 150   & 57.0 GB  & -   \\
Synthetic 30 & 357,913,900     & 150   & 113.0 GB & -   \\
Synthetic 31 & 715,827,850     & 150   & 226.0 GB & -   \\
Synthetic 32 & 1,431,655,750   & 150   & 451.0 GB & -   \\
\midrule
SRR29163078 & 10,190,262       & 151   & 3.8 GB   & \textit{P. aeruginosa}    \\
SRR28892189 & 15,137,459       & 150   & 6.3 GB   & \textit{S. coelicolor} \\
SRR26113965 & 56,271,131       & 150   & 24.0 GB  & \textit{F. vesca}   \\
SRR25743144 & 139,993,564      & 151   & 59.0 GB  & \textit{P. sinus}  \\
SRR7443702  & 141,903,420      & 125   & 45.0 GB  & \textit{Ambystoma sp.} \\
SRR28206931 & 263,469,656      & 149   & 95.0 GB  & \textit{Human}  \\
SRR29871703 & 345,818,242      & 150   & 145.0 GB & \textit{T. aestivum} \\ 
\hline    
\end{tabular}
\label{datasets}
\end{table}
\subsection{Baseline \kmer counters}

We compare against three \sota baseline implementations:
KMC3~\cite{kmc3}, PakMan~\cite{pakman}, and HySortK~\cite{hysortk}.
To make fair comparisons, we take measures to \emph{strengthen} their performance as explained below.

%\subsubsection*{3}
KMC3 is a shared memory algorithm and uses multithreaded radixsort.
It was originally designed as a disk-based out-of-core \kmer counter.
However, we use command line arguments to force KMC3 to execute in an in-memory mode, thereby yielding its \emph{best-case} performance.

%\subsubsection*{2}
PakMan's \kc kernel serves as our \MPI-only baseline.
%The first kernel of PakMan~\cite{pakman} is \kc, which we use here as the \MPI only baseline.
It communicates using a \textit{blocking} \mtom collective.
To strengthen it, we replaced its original quicksort-based \kc algorithm to use radix sort.
This change also makes it more directly comparable to KMC3, HySortK, and \akc, all of which use radix sort.
Indeed, this modification \emph{speeds up} PakMan's \kc kernel by $\approx 2\times$, as shown in \autoref{fig:pkmstar}.
We refer to this improved implementation as PakMan*.

%\subsubsection*{1} 
% HySortK uses \OMP+\MPI based hybrid parallelism.
In contrast to PakMan, HySortK uses \textit{non-blocking} \MPI \atoa collective for communication, and \OMP based multithreaded radix sorting for final counting.
On Intel nodes, we run it using different threads per MPI rank configurations and always report the \emph{best} result.
On AMD nodes, we run HySortK using one MPI rank per NUMA domain as recommended by the authors.

% \vspace{-8pt}
\subsection{Shared Memory Experiments}
\begin{figure}
    \begin{subfigure}[b]{0.45\textwidth}
        \centering
        \includesvg[width=\textwidth]{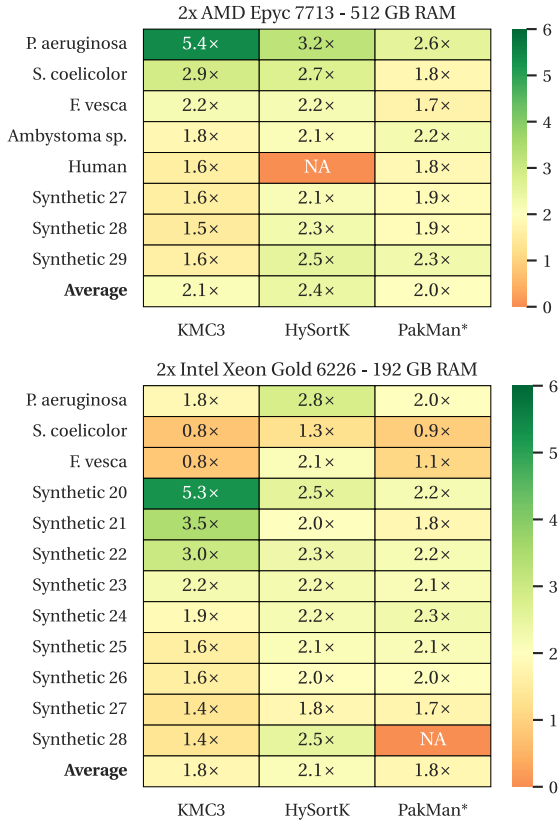}
    \end{subfigure}
    \caption{Speedup \akc over KMC3, HySortK and PakMan* on a single AMD ($128$ cores), and Intel node ($24$ cores).}
    \label{fig:sharedmem}
\end{figure}
We first ran all methods within a single shared-memory node (both AMD- and Intel-based) and summarize the results in \autoref{fig:sharedmem}.
\akc is $\approx 2\times$ faster than the other distributed memory algorithms (HySortK and PakMan*) in a shared memory environment.
Moreover, it is even $\approx 2\times$ faster than the shared memory baseline, KMC3.
This latter improvement is a collateral benefit of our choice of runtime:
the runtime detects when two PEs are colocated within a node and converts the asynchronous messages into \textsf{memcpy} calls, thereby helping us take advantage of shared memory resources without requiring that we write a separate multithreaded program~\cite{jocs}.

\subsection{Strong Scaling Experiments}
\label{experiment:strong-scaling}
We conducted strong-scaling experiments using real-world datasets and the larger synthetic datasets (scale 27 and higher), summarizing the results in \autoref{fig:strongscale} and \autoref{fig:ss32}.
Any missing data point indicates that the corresponding implementation failed due to an \OOM error;
this includes HySortK failing to count \kmers of \textit{Synthetic 32} and hence not appearing in \autoref{fig:ss32}.

All methods plateau as expected under strong scaling.
However, the best \akc configuration has a consistently lower execution time than the best configuration of the other methods.
On average, \akc is $2.34\times$ faster than HySortK, and $2.81\times$ faster than PakMan*, considering the datapoints till the strong scaling limit of \akc.

A minor artifact in these experiments is that \akc is slightly ``disadvantaged'' compared to the other methods.
By default, Conveyors decides automatically whether to run in $1D$, $2D$, or $3D$ mode.
To force it to use 1D (see Section \ref{experiment:k-D}) without modifying the library, we need to run with one \emph{fewer} core than the total available.
Thus, our implementations uses a little less concurrency, thereby indirectly strengthening the baselines.

\begin{figure}[h!]
    \centering
    \begin{subfigure}[b]{0.48\textwidth}
        \centering
        \includesvg[width=\textwidth]{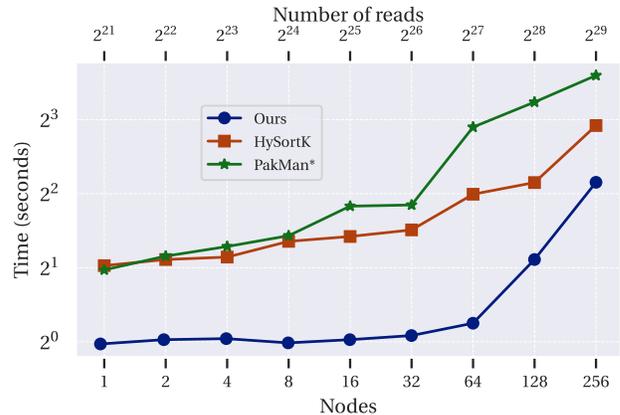}
    \end{subfigure}
    \caption{Weak scaling experiments on synthetic datasets.}
    \label{fig:weakscale}
\end{figure}
\subsection{Weak Scaling on Synthetic Datasets}

We perform weak scaling experiments on the synthetic datasets using up to $256$ Intel nodes.
\autoref{fig:weakscale} shows \akc is $1.7 - 3.4\times$, and $2.0 - 6.3\times$ faster than HySortK and PakMan* respectively.
% \autoref{fig:weakscale} shows that \akc can maintain the weak scaling efficiency till $32$ nodes / $768$ cores.
PakMan* weak-scales poorly, becoming inefficient after $2$ nodes / $48$ cores. 
HySortK weakly scales better than PakMan* but becomes inefficient after $4$ nodes / $96$ cores.
The best weak scaling efficiency is achieved by \akc, which maintains efficiency until $32$ nodes / $768$ cores.

\subsection{Blocking versus non-blocking collectives}
PakMan* and HySortK differ primarily in whether they use blocking (PakMan*) or nonblocking (HySortK) \MPI collectives.
Thus, comparing them suggests the benefit of nonblocking (ignoring performance improvements from \OMP based hybrid parallelism).
In the strong scaling experiment shown in \autoref{fig:strongscale}, HySortK is only $1.17\times$ faster than PakMan* on average.
Moreover, on $4$ out of the $7$ real datasets, (\textit{P. aeruginosa}, \textit{S. coelicolor}, \textit{F. vesca}, and \textit{Human}) PakMan*'s performance is nearly the same as HySortK.
Thus, use of nonblocking collectives does not in this case fundamentally resolve the issue of synchronization costs.

\subsection{Choice between 1D, 2D, and 3D Conveyors}
\label{experiment:k-D}
For \akc, the Conveyors runtime decides whether to use a 1D, 2D, or 3D topology automatically.
To compare them, we modified Conveyors to allow us to choose the topology and still use all available cores (unlike the \cref{experiment:strong-scaling} 1D experiments).
\autoref{fig:xdconv} shows that 1D is 10--20\% faster than $2D$ and $3D$, albeit at the cost of more memory per \autoref{fig:total_mem}.
A user in a memory constrained environment should manage this tradeoff.
\begin{figure}[h!]
    \centering
    \includesvg[width=\linewidth]{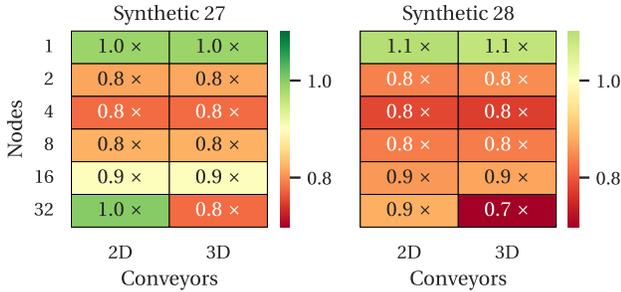}
    \caption{Speedup of 2D and 3D Conveyors over 1D.}
    \label{fig:xdconv}
\end{figure}
\subsection{Importance of aggregation protocols}
\autoref{fig:aggr} shows the benefit of incorporating application specific $L_2$ and $L_3$ protocols over the general purpose $L_0$, and $L_1$ protocols, on \textit{Human} and \textit{Synthetic 32} datasets.
To show the benefit of each of the application-specific aggregation protocols ($L_2$ and $L_3$ layers), we ran \akc with only the first two general protocols and introduced $L_2$ and $L_3$ one by one.
\textit{Synthetic 32}, is sampled from a uniform random distribution and is well-behaved by construction.
For such a dataset, the significant reduction in the number of individual messages due to $L_2$ protocol results in $\approx 2\times$ speedup over $L_0$ and $L_1$ protocols.
The overhead of $L_3$ layer does not provide any reduction in communication volume, and hence results in no additional speedup. 
\textit{Human} genome is known to have a high-frequency \kmers, and hence the $L_3$ layer is essential to achieve optimal performance.
The extra processing time in $L_3$ results in a significant reduction in communication volume, resulting in up to $66\times$ speedup over just using $L_0$ and $L_1$ aggregations. 
\begin{figure}[h!]
    \centering
    \includesvg[width=\linewidth]{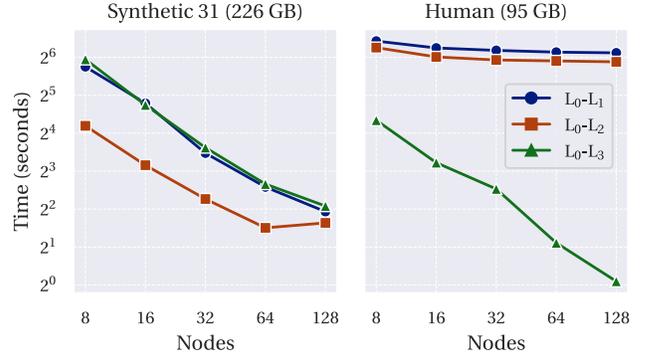}
    \caption{Strong scaling of \akc with two ($L_0 - L_1$), three ($L_0 - L_2$), and all four  ($L_0 - L_3$) aggregation protocols.}
    \label{fig:aggr}
\end{figure}
\subsection{Effects of parameter tuning}
\akc's four-layer message aggregation scheme implies four tunable parameters, and the experiments so far use the default values shown in \autoref{tab:buffers}.
When tuning the two application specific parameters, $C_2$ and $C_3$, performance is similar for $C_2 \geq 8$ but degrades for $C_2 \leq 4$, as shown in \autoref{fig:c3param}.
Similarly, per \autoref{fig:c4param} the performance remains similar for $10^3 \leq C_3 \leq 10^6$. 
Very high $C_3$ values incur additional sorting overheads, and very low $C_3$ values do not reduce reduce communication volume sufficiently.
Thus, both $C_2$ and $C_3$ should be tuned for the hardware.
\begin{figure}[h!]
    \begin{subfigure}[b]{0.48\textwidth}
        \centering
        \includesvg[width=\textwidth]{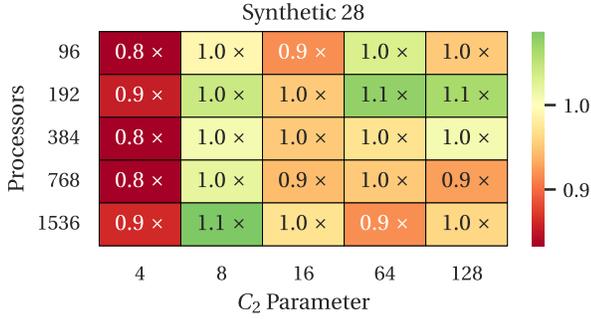}
        \caption{Different $C_2$ values over default $C_2 = 32$.}
        \label{fig:c3param}
    \end{subfigure}
    \begin{subfigure}[b]{0.43\textwidth}
        \centering
        \includesvg[width=\textwidth]{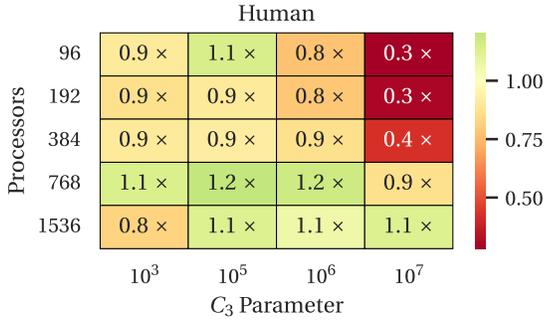}
        \caption{Different $C_3$ values over default $C_3 = 10^4$.}
        \label{fig:c4param}
    \end{subfigure}
    \caption{Tuning experiments.}
\end{figure}
%
% \subsection{Performance Breakdown}
% \textcolor{red}{\textbf{TODO:} Is it really required here? The model kinda talks about it anyway. Ask Rich, and Akihiro.}
% %
%------------------------------------------------------------
% \section{Effect on full-scale genomics pipelines} \label{genomics}
% \input{genomics}
%------------------------------------------------------------
\section{Conclusion and Future Work} \label{conclusion}
In conclusion, the \FABSP strategy of aggressive asynchrony, when combined with a carefully designed message aggregation strategy and implementation, can overcome the synchronization bottlenecks of \sota \BSP approaches for the \kc problem.

There are several avenues for future work.
Our current sorting-based approach still involves an explicit barrier between phases 1 and 2.
This synchronization could be eliminated, thereby allowing the phases to overlap, by using a distributed sorted-set data structure that supports asynchronous queries and updates.
For deployment in applications, the \kmer sizes in \akc, while sufficient for short-read genome assembly, are limited for the case of long reads due to our use of at most 64-bit integers, a limitation shared by other solutions (e.g., PakMan).
Therefore, larger integer support (e.g., 128-bit) to extend the range of supported \kmer sizes is another natural next step.
We are pursuing these directions, among others.

Another question is to what extent GPUs would benefit this workload.
Here, the answer is not clear cut.
Our analytical model suggests that memory bandwidth is one major bottleneck, so a GPU with, say, 10$\times$ more bandwidth than the CPUs used in our study would be a major win.
However, the authors of~\cite{hysortk} show that their CPU-based distributed \kmer counting significantly outperformed the GPU-based implementation of MetaHipmer2~\cite{mhmgpu}.
Indeed, our analytical model also suggests that \kc is somewhat extreme in its low operational intensity;
an estimate from our model of the op-to-byte ratio of \akc is about one 64-bit integer additions (iadd64) per 8.14 bytes or $\approx$ 0.12 iadd64/byte.
Compare this value to the much higher hardware balance of our \Phoenix CPUs of $\approx$ 2.6 iadd64/byte and an NVIDIA H100 GPU of $\approx$ 8.3 iadd64/byte.
Thus, even if a speedup is possible, the CPU units of our system are quite underutilized, and the compute units of a GPU system will be even more so.

% eof
%------------------------------------------------------------
%
\section {Acknowledgements}
This research is based upon work supported by the Office of the Director of National Intelligence (ODNI), Intelligence Advanced Research Projects Activity (IARPA), through the Advanced Graphical Intelligence Logical Computing Environment (AGILE) research program, under Army Research Office (ARO) contract number W911NF22C0083. The views and conclusions contained herein are those of the authors and should not be interpreted as necessarily representing the official policies or endorsements, either expressed or implied, of the ODNI, IARPA, or the U.S. Government.

We would like to thank Dr. Giulia Guidi for valuable discussions on computational bottlenecks in \kmer counting, and for helping us understand the HySortK algorithm and the associated software implementation.
% \newpage
\bibliography{ref}

\begin{thebibliography}{10}
\providecommand{\url}[1]{#1}
\csname url@samestyle\endcsname
\providecommand{\newblock}{\relax}
\providecommand{\bibinfo}[2]{#2}
\providecommand{\BIBentrySTDinterwordspacing}{\spaceskip=0pt\relax}
\providecommand{\BIBentryALTinterwordstretchfactor}{4}
\providecommand{\BIBentryALTinterwordspacing}{\spaceskip=\fontdimen2\font plus
\BIBentryALTinterwordstretchfactor\fontdimen3\font minus
  \fontdimen4\font\relax}
\providecommand{\BIBforeignlanguage}[2]{{%
\expandafter\ifx\csname l@#1\endcsname\relax
\typeout{** WARNING: IEEEtran.bst: No hyphenation pattern has been}%
\typeout{** loaded for the language `#1'. Using the pattern for}%
\typeout{** the default language instead.}%
\else
\language=\csname l@#1\endcsname
\fi
#2}}
\providecommand{\BIBdecl}{\relax}
\BIBdecl

\bibitem{cheng2021haplotype}
H.~Cheng, G.~T. Concepcion, X.~Feng, H.~Zhang, and H.~Li, ``Haplotype-resolved
  de novo assembly using phased assembly graphs with hifiasm,'' \emph{Nature
  Methods}, vol.~18, no.~2, pp. 170--175, 2021.

\bibitem{chikhi2014informed}
R.~Chikhi and P.~Medvedev, ``Informed and automated k-mer size selection for
  genome assembly,'' \emph{Bioinformatics}, vol.~30, no.~1, pp. 31--37, 2014.

\bibitem{koren2018novo}
S.~Koren, A.~Rhie, B.~P. Walenz, A.~T. Dilthey, D.~M. Bickhart, S.~B. Kingan,
  S.~Hiendleder, J.~L. Williams, T.~P. Smith, and A.~M. Phillippy, ``De novo
  assembly of haplotype-resolved genomes with trio binning,'' \emph{Nature
  biotechnology}, vol.~36, no.~12, pp. 1174--1182, 2018.

\bibitem{pevzner2001eulerian}
P.~A. Pevzner, H.~Tang, and M.~S. Waterman, ``An eulerian path approach to
  {DNA} fragment assembly,'' \emph{Proceedings of the national academy of
  sciences}, vol.~98, no.~17, pp. 9748--9753, 2001.

\bibitem{hipmer}
E.~Georganas, A.~Bulu{\c{c}}, J.~Chapman, S.~Hofmeyr, C.~Aluru, R.~Egan,
  L.~Oliker, D.~Rokhsar, and K.~Yelick, ``Hipmer: an extreme-scale de novo
  genome assembler,'' in \emph{Proceedings of the International Conference for
  High Performance Computing, Networking, Storage and Analysis}, 2015, pp.
  1--11.

\bibitem{pakman}
P.~Ghosh, S.~Krishnamoorthy, and A.~Kalyanaraman, ``Pakman: Scalable assembly
  of large genomes on distributed memory machines,'' in \emph{2019 IEEE
  International Parallel and Distributed Processing Symposium (IPDPS)}.\hskip
  1em plus 0.5em minus 0.4em\relax IEEE, 2019, pp. 578--589.

\bibitem{han2017concurrent}
W.~Han, M.~Wang, and Y.~Ye, ``A concurrent subtractive assembly approach for
  identification of disease associated sub-metagenomes,'' in
  \emph{International Conference on Research in Computational Molecular
  Biology}.\hskip 1em plus 0.5em minus 0.4em\relax Springer, 2017, pp. 18--33.

\bibitem{pellegrina2020fast}
L.~Pellegrina, C.~Pizzi, and F.~Vandin, ``Fast approximation of frequent k-mers
  and applications to metagenomics,'' \emph{Journal of Computational Biology},
  vol.~27, no.~4, pp. 534--549, 2020.

\bibitem{metahipmer}
S.~Hofmeyr, R.~Egan, E.~Georganas, A.~C. Copeland, R.~Riley, A.~Clum,
  E.~Eloe-Fadrosh, S.~Roux, E.~Goltsman, A.~Bulu{\c{c}} \emph{et~al.},
  ``Terabase-scale metagenome coassembly with metahipmer,'' \emph{Scientific
  reports}, vol.~10, no.~1, p. 10689, 2020.

\bibitem{mhmgpu}
M.~G. Awan, S.~Hofmeyr, R.~Egan, N.~Ding, A.~Buluc, J.~Deslippe, L.~Oliker, and
  K.~Yelick, ``Accelerating large scale de novo metagenome assembly using
  gpus,'' in \emph{Proceedings of the International Conference for High
  Performance Computing, Networking, Storage and Analysis}, 2021, pp. 1--11.

\bibitem{rhie2020merqury}
A.~Rhie, B.~P. Walenz, S.~Koren, and A.~M. Phillippy, ``Merqury: reference-free
  quality, completeness, and phasing assessment for genome assemblies,''
  \emph{Genome biology}, vol.~21, no.~1, pp. 1--27, 2020.

\bibitem{kelley2010quake}
D.~R. Kelley, M.~C. Schatz, and S.~L. Salzberg, ``Quake: quality-aware
  detection and correction of sequencing errors,'' \emph{Genome biology},
  vol.~11, no.~11, pp. 1--13, 2010.

\bibitem{salmela2017accurate}
L.~Salmela, R.~Walve, E.~Rivals, and E.~Ukkonen, ``Accurate self-correction of
  errors in long reads using de bruijn graphs,'' \emph{Bioinformatics},
  vol.~33, no.~6, pp. 799--806, 2017.

\bibitem{li2003estimating}
X.~Li and M.~S. Waterman, ``Estimating the repeat structure and length of {DNA}
  sequences using l-tuples,'' \emph{Genome research}, vol.~13, no.~8, pp.
  1916--1922, 2003.

\bibitem{audoux2017kupl}
J.~Audoux, N.~Philippe, R.~Chikhi, M.~Salson, M.~Gallopin, M.~Gabriel,
  J.~Le~Coz, E.~Drouineau, T.~Commes, and D.~Gautheret, ``De-kupl: exhaustive
  capture of biological variation in {RNA}-seq data through k-mer
  decomposition,'' \emph{Genome biology}, vol.~18, no.~1, pp. 1--15, 2017.

\bibitem{chong2017novobreak}
Z.~Chong, J.~Ruan, M.~Gao, W.~Zhou, T.~Chen, X.~Fan, L.~Ding, A.~Y. Lee,
  P.~Boutros, J.~Chen \emph{et~al.}, ``novobreak: local assembly for breakpoint
  detection in cancer genomes,'' \emph{Nature methods}, vol.~14, no.~1, pp.
  65--67, 2017.

\bibitem{khorsand2021nebula}
P.~Khorsand and F.~Hormozdiari, ``Nebula: Ultra-efficient mapping-free
  structural variant genotyper,'' \emph{Nucleic acids research}, vol.~49,
  no.~8, pp. e47--e47, 2021.

\bibitem{kmc2}
S.~Deorowicz, M.~Kokot, S.~Grabowski, and A.~Debudaj-Grabysz, ``{KMC} 2: Fast
  and resource-frugal k-mer counting,'' \emph{Bioinformatics}, vol.~31, no.~10,
  pp. 1569--1576, 2015.

\bibitem{kurtz_new_2008}
S.~Kurtz, A.~Narechania, J.~C. Stein, and D.~Ware, ``A new method to compute
  k-mer frequencies and its application to annotate large repetitive plant
  genomes,'' \emph{BMC Genomics}, vol.~9, no.~1, p. 517, 2008.

\bibitem{li_mspkmercounter:_2015}
Y.~Li \emph{et~al.}, ``{MSPKmerCounter}: a fast and memory efficient approach
  for k-mer counting,'' \emph{arXiv preprint arXiv:1505.06550}, 2015.

\bibitem{li2018gpu}
H.~Li, A.~Ramachandran, and D.~Chen, ``{GPU} acceleration of advanced k-mer
  counting for computational genomics,'' in \emph{2018 IEEE 29th International
  Conference on Application-specific Systems, Architectures and Processors
  (ASAP)}.\hskip 1em plus 0.5em minus 0.4em\relax IEEE, 2018, pp. 1--4.

\bibitem{mamun_kcmbt_2016}
A.-A. Mamun, S.~Pal, and S.~Rajasekaran, ``Kcmbt: a k-mer counter based on
  multiple burst trees,'' \emph{Bioinformatics}, vol.~32, no.~18, pp.
  2783--2790, 2016.

\bibitem{marcais_fast_2011}
G.~Mar{\c{c}}ais and C.~Kingsford, ``A fast, lock-free approach for efficient
  parallel counting of occurrences of k-mers,'' \emph{Bioinformatics}, vol.~27,
  no.~6, pp. 764--770, 2011.

\bibitem{pan2018optimizing}
T.~C. Pan, S.~Misra, and S.~Aluru, ``Optimizing high performance distributed
  memory parallel hash tables for {DNA} k-mer counting,'' in \emph{SC18:
  International Conference for High Performance Computing, Networking, Storage
  and Analysis}.\hskip 1em plus 0.5em minus 0.4em\relax IEEE, 2018, pp.
  135--147.

\bibitem{pandey2018squeakr}
P.~Pandey, M.~A. Bender, R.~Johnson, and R.~Patro, ``Squeakr: an exact and
  approximate k-mer counting system,'' \emph{Bioinformatics}, vol.~34, no.~4,
  pp. 568--575, 2018.

\bibitem{rizk_dsk:_2013}
G.~Rizk, D.~Lavenier, and R.~Chikhi, ``{DSK}: k-mer counting with very low
  memory usage,'' \emph{Bioinformatics}, vol.~29, no.~5, pp. 652--653, 2013.

\bibitem{kmc3}
M.~Kokot, M.~D{\l}ugosz, and S.~Deorowicz, ``Kmc 3: counting and manipulating
  k-mer statistics,'' \emph{Bioinformatics}, vol.~33, no.~17, pp. 2759--2761,
  2017.

\bibitem{hysortk}
Y.~Li and G.~Guidi, ``High-performance sorting-based k-mer counting in
  distributed memory with flexible hybrid parallelism,'' in \emph{Proceedings
  of the 53rd International Conference on Parallel Processing}, 2024, pp.
  919--928.

\bibitem{kmerindsc18}
T.~C. Pan, S.~Misra, and S.~Aluru, ``Optimizing high performance distributed
  memory parallel hash tables for dna k-mer counting,'' in \emph{SC18:
  International Conference for High Performance Computing, Networking, Storage
  and Analysis}.\hskip 1em plus 0.5em minus 0.4em\relax IEEE, 2018, pp.
  135--147.

\bibitem{jocs}
\BIBentryALTinterwordspacing
S.~R. Paul, A.~Hayashi, K.~Chen, Y.~Elmougy, and V.~Sarkar, ``A fine-grained
  asynchronous bulk synchronous parallelism model for pgas applications,''
  \emph{Journal of Computational Science}, vol.~69, p. 102014, 2023. [Online].
  Available:
  \url{https://www.sciencedirect.com/science/article/pii/S1877750323000741}
\BIBentrySTDinterwordspacing

\bibitem{kmc1}
S.~Deorowicz, A.~Debudaj-Grabysz, and S.~Grabowski, ``Disk-based k-mer counting
  on a pc,'' \emph{BMC bioinformatics}, vol.~14, pp. 1--12, 2013.

\bibitem{marccais2017improving}
G.~Mar{\c{c}}ais, D.~Pellow, D.~Bork, Y.~Orenstein, R.~Shamir, and
  C.~Kingsford, ``Improving the performance of minimizers and winnowing
  schemes,'' \emph{Bioinformatics}, vol.~33, no.~14, pp. i110--i117, 2017.

\bibitem{raduls}
M.~Kokot, S.~Deorowicz, and A.~Debudaj-Grabysz, ``Sorting data on ultra-large
  scale with raduls,'' in \emph{International Conference: Beyond Databases,
  Architectures and Structures}.\hskip 1em plus 0.5em minus 0.4em\relax
  Springer, 2017, pp. 235--245.

\bibitem{raduls2}
M.~Kokot, S.~Deorowicz, and M.~D{\l}ugosz, ``Even faster sorting of (not only)
  integers,'' in \emph{Man-Machine Interactions 5: 5th International Conference
  on Man-Machine Interactions, ICMMI 2017 Held at Krak{\'o}w, Poland, October
  3-6, 2017}.\hskip 1em plus 0.5em minus 0.4em\relax Springer, 2018, pp.
  481--491.

\bibitem{melsted_efficient_2011}
P.~Melsted and J.~K. Pritchard, ``Efficient counting of k-mers in {DNA}
  sequences using a bloom filter,'' \emph{BMC Bioinformatics}, vol.~12, no.~1,
  p.~1, 2011.

\bibitem{cqf_pandey}
P.~Pandey, M.~A. Bender, R.~Johnson, and R.~Patro, ``A general-purpose counting
  filter: Making every bit count,'' in \emph{Proceedings of the 2017 ACM
  International Conference on Management of Data}, ser. SIGMOD '17.\hskip 1em
  plus 0.5em minus 0.4em\relax New York, NY, USA: Association for Computing
  Machinery, 2017, p. 775–787.

\bibitem{manekar2018benchmark}
S.~C. Manekar and S.~R. Sathe, ``A benchmark study of k-mer counting methods
  for high-throughput sequencing,'' \emph{GigaScience}, vol.~7, no.~12, p.
  giy125, 2018.

\bibitem{mccoy2023singleton}
H.~McCoy, S.~Hofmey, K.~Yelick, and P.~Pandey, ``Singleton sieving: Overcoming
  the memory/speed trade-off in exascale $\kappa$-mer analysis,'' in \emph{SIAM
  Conference on Applied and Computational Discrete Algorithms (ACDA23)}.\hskip
  1em plus 0.5em minus 0.4em\relax SIAM, 2023, pp. 213--224.

\bibitem{nisa2021distributed}
I.~Nisa, P.~Pandey, M.~Ellis, L.~Oliker, A.~Bulu{\c{c}}, and K.~Yelick,
  ``Distributed-memory k-mer counting on gpus,'' in \emph{2021 IEEE
  International Parallel and Distributed Processing Symposium (IPDPS)}.\hskip
  1em plus 0.5em minus 0.4em\relax IEEE, 2021, pp. 527--536.

\bibitem{cheng2024rapidgkc}
Y.~Cheng, X.~Sun, and Q.~Luo, ``Rapidgkc: Gpu-accelerated k-mer counting,'' in
  \emph{2024 IEEE 40th International Conference on Data Engineering
  (ICDE)}.\hskip 1em plus 0.5em minus 0.4em\relax IEEE, 2024, pp. 3810--3822.

\bibitem{georganas2014parallel}
E.~Georganas, A.~Bulu{\c{c}}, J.~Chapman, L.~Oliker, D.~Rokhsar, and K.~Yelick,
  ``Parallel de bruijn graph construction and traversal for de novo genome
  assembly,'' in \emph{SC'14: Proceedings of the International Conference for
  High Performance Computing, Networking, Storage and Analysis}.\hskip 1em plus
  0.5em minus 0.4em\relax IEEE, 2014, pp. 437--448.

\bibitem{elba}
G.~Guidi, O.~Selvitopi, M.~Ellis, L.~Oliker, K.~Yelick, and A.~Bulu{\c{c}},
  ``Parallel string graph construction and transitive reduction for de novo
  genome assembly,'' in \emph{2021 IEEE International Parallel and Distributed
  Processing Symposium (IPDPS)}.\hskip 1em plus 0.5em minus 0.4em\relax IEEE,
  2021, pp. 517--526.

\bibitem{KmerInd}
T.~Pan, P.~Flick, C.~Jain, Y.~Liu, and S.~Aluru, ``Kmerind: A flexible parallel
  library for k-mer indexing of biological sequences on distributed memory
  systems,'' in \emph{Proceedings of the 7th ACM international conference on
  bioinformatics, computational biology, and health informatics}, 2016, pp.
  422--433.

\bibitem{mpi}
M.~P. Forum, ``Mpi: A message-passing interface standard,'' USA, Tech. Rep.,
  1994.

\bibitem{openshmem}
B.~Chapman, T.~Curtis, S.~Pophale, S.~Poole, J.~Kuehn, C.~Koelbel, and
  L.~Smith, ``Introducing openshmem: Shmem for the pgas community,'' in
  \emph{Proceedings of the Fourth Conference on Partitioned Global Address
  Space Programming Model}, 2010, pp. 1--3.

\bibitem{conv}
F.~M. Maley and J.~G. DeVinney, ``Conveyors for streaming many-to-many
  communication,'' in \emph{2019 IEEE/ACM 9th Workshop on Irregular
  Applications: Architectures and Algorithms (IA3)}, 2019, pp. 1--8.

\bibitem{skasort}
M.~Skarupke, ``I wrote a faster sorting algorithm,''
  \url{https://probablydance.com/2016/12/27/i-wrote-a-faster-sorting-algorithm/},
  2017, accessed: 2024-09-18.

\bibitem{papi}
P.~J. Mucci, S.~Browne, C.~Deane, and G.~Ho, ``Papi: A portable interface to
  hardware performance counters,'' in \emph{Proceedings of the department of
  defense HPCMP users group conference}, vol. 710, 1999.

\bibitem{huang2012art}
W.~Huang, L.~Li, J.~R. Myers, and G.~T. Marth, ``Art: a next-generation
  sequencing read simulator,'' \emph{Bioinformatics}, vol.~28, no.~4, pp.
  593--594, 2012.

\bibitem{ncbi}
E.~W. Sayers, E.~E. Bolton, J.~R. Brister, K.~Canese, J.~Chan, D.~C. Comeau,
  R.~Connor, K.~Funk, C.~Kelly, S.~Kim \emph{et~al.}, ``Database resources of
  the national center for biotechnology information,'' \emph{Nucleic acids
  research}, vol.~50, no.~D1, pp. D20--D26, 2022.

\bibitem{sratoolkit}
S.~T.~D. Team, ``The ncbi sra toolkit github,''
  \url{https://github.com/ncbi/sra-tools}, 2024, accessed: 2024-10-02.

\end{thebibliography}
% \appendix
\end{document}

\appendix
\label{hclib1Dselectseeds}
\begin{figure}[hbp]
    \centering
    \begin{subfigure}[b]{0.45\textwidth}
        \centering
        \includegraphics[width=\textwidth]{figures/cartoon5.png}
    \end{subfigure}
    \caption{HClib \texttt{SelectSeeds} 1D Implementation}
\end{figure}